\documentclass{llncs}

\usepackage{inputenc}
\usepackage{indentfirst}
\usepackage{longtable}
\usepackage{amstext, amssymb, amsmath}
\begin{document}

\title{Undecidability of performance equivalence of Petri nets
%\thanks{% Work reported here has been
%        This work was supported by the
%        EU Research Training Network {\sc Games}.}
}
%\thanks{% Work reported here has been
%        This work is supported by the
%        European Community Research Training Network {\sc Games}.}
%
\author{S{\l}awomir Lasota\thanks{The first author acknowledges a partial support of the National Science Centre grant 2013/09/B/ST6/01575.}
%Polish government grant no. N206 008 32/0810.}
\and Marcin Poturalski}
\institute{Institute of Informatics, Warsaw University\\
\email{sl@mimuw.edu.pl} \and
EPFL, Lausanne, Switzerland\\
\email{marcin.poturalski@epfl.ch}}

\maketitle

%\runninghead{S. Lasota, M. Poturalski}{Undecidability of performance equivalence of Petri nets}

\begin{abstract}
We investigate bisimulation equivalence on Petri nets under durational semantics.
Our motivation was to verify the conjecture that in durational setting,
the bisimulation equivalence checking problem becomes more tractable than in ordinary setting (which is the case,
e.g., over communication-free nets).
We disprove this conjecture in three of four proposed variants of durational semantics.
The fourth variant remains an intriguing open problem.
\end{abstract}

%\begin{keywords}
%Petri nets, bisimulation equivalence, performance equivalence
%\end{keywords}

%\newtheorem{definition}{Definition}%[section]
%\newtheorem{theorem}{Theorem}
%\newtheorem{fact}[theorem]{Fact}
%\newtheorem{corollary}[theorem]{Corollary}
%\newtheorem{example}{Example}

%\newcommand{\noproof}{\par \hspace{0.9\hsize} $\Box$\par}
%\newenvironment{dw}{\par {\setlength{\parindent}{0in} \textbf{Dowód:}} \par}{\bezdw}

\newcommand{\patient}{patient}
\newcommand{\Patient}{Patient}
\newcommand{\impatient}{impatient}
\newcommand{\Impatient}{Impatient}
\newcommand{\localtime}{local-time}
\newcommand{\Localtime}{Local-time}
\newcommand{\globaltime}{global-time}
\newcommand{\Globaltime}{Global-time}
\newcommand{\ourpr}{performance equivalence}
\newcommand{\Ourpr}{Performance equivalence}
\newcommand{\ourpradj}{performance equivalent}
\newcommand{\equi}{equimarking} % equipoint? equimarking?
\newcommand{\Equi}{Equimarking}
\newcommand{\A}{$\mathtt{Spoi\-ler}$}
\newcommand{\D}{$\mathtt{Dup\-li\-ca\-tor}$}
\newcommand{\M}{\mathbb{M}}

\newcommand{\tssize}{0.55\hsize}

\newcommand{\defi}[1]{\textsl{#1}}
\newcommand{\multi}[1]{#1^*}
\newcommand{\pre}[1]{{}^{{}^{{}_\bullet}}\!#1}
\newcommand{\post}[1]{#1^{{}^{{}_\bullet}}}
\newcommand{\mrk}[1]{\mathsf{M}(#1)}
\newcommand{\U}[1]{\mathsf{U}(#1)}
\newcommand{\atrans}[1]{\stackrel{#1}{\longrightarrow}}
\newcommand{\trans}[1]{\xrightarrow{#1}}
\newcommand{\dtrans}[1]{\downarrow^{{}_{#1}}}
\newcommand{\ddtrans}[2]{#2\cdot\!\downdownarrows^{{}_{#1}}}
\newcommand{\wtrans}[1]{\stackrel{#1}{\Longrightarrow}}
\newcommand{\tr}{\triangleright}
\newcommand{\comp}{\parallel}

\newcommand{\typeInc}{(increment)}
\newcommand{\typeIf}{(zero-test or decrement)}
\newcommand{\typeHalt}{(halting instruction)}

\newcommand{\instrInc}{\begin{quote}
$\mathtt{i: c_b := c_b + 1;\ goto\ j}$ \hspace{2.5cm} \typeInc
\end{quote}
}
\newcommand{\instrIf}{\begin{quote}
\begin{tabular}{@{}l@{}l}
$\mathtt{i: if\ c_b = 0\ }$ & $\mathtt{then\ goto\ k}$ \hspace{2.2cm} \typeIf \\
                            & $\mathtt{else\ c_b := c_b - 1;\ goto\ j}$ \\
\end{tabular}
\end{quote}
}
\newcommand{\instrHalt}{\begin{quote}
$\mathtt{n: halt}$ \hspace{5cm} \typeHalt
\end{quote}
}

\newcommand{\calR}{{\cal R}}
\newcommand{\calS}{{\cal S}}
\newcommand{\calA}{{\cal A}}
\newcommand{\calL}{{\cal L}}
\newcommand{\calP}{{\cal P}}
\newcommand{\bisim}{\sim}
\newcommand{\nat}{\mathbb{N}}
\newcommand{\tranrule}[1]{\stackrel{#1}{\leadsto}}
\newcommand{\dur}[1]{\mathtt{dur}(#1)}
\newcommand{\timed}{durational}
\newcommand{\Timed}{Durational}
\newcommand{\untime}[1]{\mathtt{untime}(#1)}
\newcommand{\maxstamp}[1]{\mathtt{max\text{-}stamp}(#1)}
\newcommand{\stamps}[1]{\mathtt{stamps}(#1)}
\newcommand{\untiming}{untiming}
\newcommand{\removed}[1]{}

\section{Introduction}

Bisimulation equivalence~\cite{park,mil} is one of the most relevant semantical equivalences
of concurrent systems. One of its advantages is that it often allows for an
efficient verification algorithms in settings where other approaches (see, e.g., \cite{vgl}), like
language equality, lead to undecidable verification problems.
There is now a wide range of results about decidability and complexity of
different variants of bisimulation equivalences in different classes of infinite-state
systems (see e.g.~\cite{bcms}).

Bisimulation equivalence relates processes exhibiting the same behaviour.
In this paper we investigate \emph{performance equivalence}, a variant of
bisimulation equivalence that aims at relating not only purely functional behaviour, but
also effectiveness of processes.
A basic assumption is that each action of a process has assigned a positive \emph{duration},
that is amount of time (or other resource) necessary to complete this action.
Performance equivalence is then a variant of bisimulation that respects
amount of time (resource) requested in both processes during execution.
This notion was introduced in~\cite{grs} and then studied
among the others in~\cite{c,cgr,bls,las-concur,las-tcs}.

A starting point for our investigations was an observation made in~\cite{bls}
that the complexity of performance equivalence may be substantially lower than
complexity of ordinary bisimulation equivalence.
The authors of~\cite{bls} investigated so called Basic Parallel Processes (BPP in short),
a natural and simple fragment of process algebra CCS~\cite{mil} (it is expressibly equivalent
to CCS without communication).
BPP, when transformed to a normal form~\cite{chr}, is equivalent to
communication-free Petri nets.
While bisimulation equivalence on BPP in normal form is PSPACE-complete~\cite{janbpp,srb},
in~\cite{bls} it was shown that performance equivalence may be decided in polynomial time.
An intuitive justification of this is that the latter equivalence, being more discriminating,
satisfies stronger decomposition properties and hence is more tractable.
Later on, it was shown that the polynomial time procedure exists for the whole BPP
(not necessarily in the normal form studied in~\cite{bls})
and that it coincides with distributed equivalence~\cite{las-tcs}.

Performance equivalence is computationally more tractable than
bisimulation equivalence on BPP, i.e., on communication-free Petri nets,
hence a natural question arises: is it also more tractable in the case of general
Petri nets?
As bisimulation equivalence is undecidable in this case~\cite{jan},
the crucial question is whether performance equivalence is decidable or not.
This is the main problem investigated in this paper.

However, when one tries to define the \timed\ semantics over Petri nets,
necessary to host the notion of performance equivalence, it quickly turns out that
there is no unique such semantics. We made a systematic research of possible ways to define
it and come up with four different variants of \timed\ semantics.
The distinction depends on the choice between \globaltime\ or \localtime\ approach,
and on the way of synchronisation (\patient\ and \impatient\ approach).

As our main result, we proved undecidability of performance equivalence under
three of the four semantics.
While the \patient\ variants are easily undecidable, the proof for
\globaltime\ \impatient\ semantics is nontrivial and
constitutes the main technical contribution of the paper.
The proof builds on the method of Jan\v{c}ar~\cite{jan};
however, substantial new insight was necessary to come over new difficulties
and subtleties appearing in the durational setting.

Under the \impatient\ \localtime\ semantics, the question is still open.
If decidable, performance equivalence would be one of very few notions of equivalence
of general Petri nets exhibiting a decision procedure.
This is actually the main motivation of this paper:

\begin{quote}
{\bf Motivation:}\ basing on the positive impact of durational semantics on complexity
of equivalence-checking for BPP-nets, attempt to prove dedidability for general nets,
thus discovering a decidable bisimulation-like equivalence of general nets.
\end{quote}

There is a wide range of research on timed extensions of Petri nets,
like time nets or timed nets~\cite{merlin,ramcham}.
In most of these extensions, some timing restrictions are posed on transition,
places, or arcs.
However, we would like to stress that our \timed\ setting is different from the timed ones.
The principal difference is that we do not aim at modelling timed behaviour;
instead, our aim is to \emph{measure} effectiveness of processes.
In particular, we allow for a \localtime\ semantics, where the time-stamps
observed during an execution of a net need not be a monotonic sequence.

Another distinguishing aspect is that the timed settings usually properly extend
ordinary untimed nets, therefore decision problems become never easier,
and typically harder.
As a relevant example, the reachability problem, decidable for ordinary nets,
becomes undecidable both for time nets and timed nets~\cite{nondecid}.

Our \timed\ setting \emph{does not} subsume ordinary nets.
This gives hope for decidability of performance equivalence, and this
stands behind the fact that the reachability problem is decidable in all four
variants (in contrast to time nets and timed nets).
This topic is discussed in detail in the last section.
The \timed\ setting does properly extend ordinary nets only when we allow
for zero durations; see~\cite{las-concur} for decidability results
about performance equivalence over BPP.

In Section~\ref{s:prel} we introduce the background material and formalise
the durational setting. We also mention quickly undecidability in both
\patient\ variants.
Then Section~\ref{s:proof} is devoted to the proof of undecidability under
\globaltime\ \impatient\ variant.
The last section contains a brief discussion on decidability of
reachability problem for \timed\ Petri nets and some final remarks.

\section{Preliminaries}
\label{s:prel}

We start with the necessary background and notation (Sections~\ref{sec:bisim} and~\ref{sec:pn}), and 
then in Section~\ref{sec:durpn} we define durational Petri nets, the model investigated in this paper.
Finally, in Sections~\ref{sec:warmup1} and~\ref{sec:warmup2} we discuss some easy cases, 
as a warmup before the main technical result to be presented in the next section.

\subsection{\bf Bisimulation equivalence}  \label{sec:bisim}
A labelled transition system consists of a set of states $\calS$ and a family of
binary relations $\{ \trans{a} \}_{a \in \calL}$ indexed by a labelling set $\calL$.
We will write $s_1 \trans{a} s_2$ instead of $(s_1, s_2) \in \trans{a}$.

A relation $\calR\subseteq \calS\times \calS$ is a \emph{bisimulation} if
for each $(s_1,s_2)\in\calR$ the following two conditions hold:
\begin{itemize}
\item if $s_1\trans{a}t_1$ for some $a,t_1$ then there is some
  $t_2$ such that $s_2 \trans{a} t_2$ and $(t_1,t_2)\in\calR$;
\item if $s_2\trans{a}t_2$ for some $a,t_2$ then there is some
  $t_1$ such that $s_1\trans{a} t_1$ and $(t_1,t_2)\in\calR$.
\end{itemize}
States $s_1,s_2$ are \emph{bisimulation equivalent} (\emph{bisimilar}),
% written $s_1 \bisim s_2$,
if there is a bisimulation $\calR$ containing
$(s_1,s_2)$.
% The relation $\bisim$ is called \emph{bisimulation equivalence} or \emph{bisimilarity}.

It is instructive to recall an alternative definition of bisimilarity,
in a setting of games.
A Bisimulation Game is played between two players \A\ and \D.
For convenience of presentation, we view \A\ as ''him'' and \D\ as ''her''.
The \emph{positions} in the game are pairs $(s_1,s_2)\in \calS\times\calS$.
In a position $(s_1,s_2)$, \A\ chooses $i\in\{1,2\}$ and
a transition from $s_i$, say $s_i\trans{a} t_i$;
\D\ must respond by choosing some transition
with the same label $a$ from the other element of $(s_1,s_2)$, i.e.,
a transition $s_{3-i}\trans{a} t_{3-i}$\,.
%($3-i=2$ when $i=1$, and  $3-i=1$ when $i=2$).
The play then continues from the position $(t_1,t_2)$.
If one of the players gets stuck (there is no
appropriate transition), the other player wins.
If the play continues forever, \D\ wins unconditionally.

Generally speaking, a \emph{strategy} for a player $P$
in a game is a (partial) function which determines a concrete $P$-move
for each sequence of so far played moves after which it is $P$'s turn.
A strategy is \emph{winning} if player $P$ wins each play
when he/she uses the strategy.
In what follows, by a strategy we always mean a \emph{memoryless strategy}:
each prescribed move depends on the current position only,
not on the whole sequence of so-far played moves.
By standard results, for each position $(s_1, s_2)$, precisely one of the players
has a memoryless winning strategy; moreover we have:
\begin{proposition}[\cite{Stirling:1997}]
Two states $s_1$ and $s_2$ are bisimilar iff \D\ has a winning strategy
in Bisimulation Game starting from position $(s_1, s_2)$.
\end{proposition}
Hence \A\ has a winning strategy iff $s_1$ and $s_2$ are \emph{not} bisimilar.

\subsection{\bf Labelled Petri nets} \label{sec:pn}
A finite multiset over a set $\calA$ is formally a mapping $M$
from $\calA$ to $\nat$, the set of natural numbers, such that $M(a) > 0$ for only
a finite number of elements $a$.
E.g., the empty multiset $\emptyset$ maps all $a \in \calA$ to $0$.
We apply the usual arithmetical operations to multisets in a point-wise manner.
E.g., point-wise addition is the union operation of multisets; and $M \leq M'$ means
that $M(a) \leq M'(a)$ for each $a \in \calA$.
We will often write finite multisets by enumerating its elements, e.g.,
$aab$ will denote the function mapping $a$ to $2$, $b$ to $1$, and all other elements
of $\calA$ to $0$.

A \emph{labelled Petri net} (\emph{net} in short)
$N$ is given by a finite set of places $\calP$, a finite
set $\calL$ of labels, and a finite set of
transition rules of the form $X \tranrule{a} Y$, where $X$ and $Y$ are finite nonempty
multisets over $\calP$ and $a \in \calL$. 
% Each of $X$, $Y$ may be possibly empty.
In the sequel it will be sufficient to consider only those nets, where
$X$ and $Y$ are always sets (i.e., multisets mapping each element to $0$ or $1$).
This restriction corresponds to the class of pure Place/Transition labelled Petri nets.

A net naturally induces a labelled transition system. Its states are all
finite multisets over $\calP$ (traditionally called \emph{markings}).
There is a transition from marking $M$ to $M'$, labelled by $a$,
if there is a transition rule $X \tranrule{a} Y$ such that
$M \geq X$ (point-wise) and $M' = M - X + Y$.
In traditional terminology, one says that the transition rule $X \tranrule{a} Y$ is
\emph{fired} at marking $M$; the transition rule is \emph{fireable} in $M$ if
$M \geq X$.
Note that we do not fix an \emph{initial marking} of a net.

There is a natural operational interpretation of the induced transition system.
A marking may be understood as an assignment of a number of \emph{tokens} to each place.
And each transition rule specifies the necessary condition on the number of tokens on some
places.
For instance, a rule $ppq \tranrule{a} ps$ would mean that an $a$-transition is allowed
assumed at least two tokens on place $p$ and at least one on place $q$;
as an outcome of the transition, the three tokens are removed, then
one token is placed back on $p$ and one on $s$.

The bisimulation equivalence problem for labelled Petri nets is defined as follows:
given a net $N$ and two markings $M_1, M_2$, decide whether they are bisimilar
as the states in the labelled transition system induced by $N$.
Undecidability of this problem was proved by Jan\v{c}ar in~\cite{jan}.
Note that equivalently we could formulate the problem for the (initial) markings in two distinct 
nets.

Bisimulation equivalence relates two states (or two systems, in general)
with the same functional behaviour.
We will now extend this setting to allow to compare, in addition to pure functional
behaviour, also its effectiveness (performance).
The idea comes from~\cite{grs} and amounts to (1) assigning a \emph{duration} to each transition;
(2) respecting the durations in the bisimulation equivalence.

\subsection{\bf \Timed\ labelled Petri nets} \label{sec:durpn}
Following~\cite{grs}, we choose a discrete time domain, represented by natural numbers.
From now on
we will assume that each transition rule $r$ of any labelled Petri net $N$ has assigned a positive
natural number, its \emph{duration}, written $\dur{r}$.
Such nets, enriched by a duration function, are called \emph{\timed\ nets} in the sequel.
Note that we do not allow for $\dur{r} = 0$, as in~\cite{grs} and in the following 
papers~\cite{c,cgr,bls}.
If we allowed for zero durations our model would trivially subsume ordinary (non-\timed) Petri nets.

A \emph{\timed\ marking} is a finite multiset over $\calP \times \nat$;
intuitively, each token in a marking has now a \emph{time-stamp}.
In an initial marking, the time-stamps of all the tokens will be usually $0$.
We will write $t \tr p$ instead of $(p, t)$.
A \timed\ marking $M$ may be naturally mapped to an ordinary marking $\untime{M}$
by removing all the time-stamps (\untiming) but preserving the number of tokens at each place.
E.g., the \timed\ marking $M = (0 \tr p) (3 \tr p) (3 \tr p) (2 \tr q)$ would be mapped to
$\untime{M} = pppq$.

For notational convenience we will extend the $t \tr \_$ notation to non-\timed\ markings: by $t \tr M$ 
we will mean the multiset $\{ t \tr p : p \in M \}$. 
E.g., $(0 \tr p) (3 \tr p) (3 \tr p) (2 \tr q)$ may be equivalently written as 
$(0 \tr p) (3 \tr pp) (2 \tr q)$.

We will define the \timed\ semantics for Petri nets by specifying
(i) when a transition rule is fireable in a \timed\ marking $M$,
and (ii) what is the effect of its firing.
The labels in the induced transition system will be now pairs $(a, t) \in \calL\times\nat$.
The intuitive meaning of a transition $M \trans{a,t} M'$ is that
$t$ units of time had to elapse \emph{before} this transition became fireable.
The amount of time $t$ is always measured relative to the starting moment of an execution;
in particular, the very first transition will be usually fired at $t = 0$.

Before making our \timed\ semantics explicit, we need to introduce some notation.
For a \timed\ marking $M$, by $\stamps{M}$ denote the set (not multiset)
of all time-stamps of tokens
in $M$, and by $\maxstamp{M}$ the greatest time-stamp in $M$.
Formally:
\[
\stamps{M} = \{ t : t \tr p \in M \} \ \ \ \ \ \ \ \ \
\maxstamp{M} = \begin{cases}
\max( \stamps{M} ) & \text{if\ } M \neq \emptyset \\
0 & \text{otherwise}
\end{cases}
\]
We distinguish four different \timed\ semantics, depending on whether it is
\patient\ or \impatient, and whether it is \globaltime\ or \localtime.
In each of the four variants, we will make it precise when \emph{a transition rule
$X \tranrule{a} Y$ is fireable in a \timed\ marking $M$, at time $t$ due to a submarking $\bar{X}$}.
In each of the four variants, it requires that $\bar{X} \leq M$ ($\bar{X}$ is a
\emph{\timed\ submarking} of $M$) and $X = \untime{\bar{X}}$ (the rule
actually applies to $\bar{X}$).
In addition, the following is required, depending on the variant of semantics:
\begin{itemize}
\item \localtime, \patient\ semantics: $\maxstamp{\bar{X}} = t$.
\item \localtime, \impatient\ semantics: $\stamps{\bar{X}} = \{t\}$.
\end{itemize}
Intuitively, in the \patient\ semantics, a token with an ''earlier'' (smaller) time-stamp may
''wait'' for other tokens with ''later'' time-stamps; but this is not allowed
in the \impatient\ variant, where the tokens must agree on their time-stamps to
be able to fire synchronously a transition rule.

Note that we did not assume that a transition with the smallest possible time-stamp
is chosen to be fired. In the \globaltime\
\patient\ (impatient) semantics we additionally require this:
a transition rule $X \tranrule{a} Y$ is fireable in $M$,
at time $t$ due to $\bar{X}$, if it is fireable according to the
\localtime\ \patient\ (\impatient) semantics, and $t$ is the smallest among all
possible choices of a rule $X \tranrule{a} Y$ and a submarking $\bar{X}$.

Uniformly for all four variants, whenever a transition rule $r = (X \tranrule{a} Y)$
is fireable in $M$, at time $t$ due to $\bar{X}$,
then in the induced labelled transition system there is a transition
$M \trans{a, t} M'$, where $M' = M - \bar{X} + \{ (t+\dur{r}) \tr p : p \in Y \}$.
I.e., fresh tokens are produced as specified by $Y$,
and their time-stamps are equal to $t + \dur{r}$.
We call $t$ a \emph{time label} of transition $M \trans{a,t} M'$.

When convenient, we will identify a Petri net with its induced transition system,
e.g., we will speak of 'transitions' of a net.
Note that the induced transition system is always acyclic, due to the assumption that durations are strictly positive.
By an execution of a net we mean a finite or infinite sequence of transitions:
\[
M_0 \trans{a_0, t_0} M_1 \trans{a_1, t_1} M_2 \trans{a_2, t_2} \ldots
\]
Note that the initial marking $M_0$ is not fixed for the net.
For two \timed\ markings $M$, $M'$, we say that $M'$ is \emph{reachable} from $M$
if there is a finite execution that starts in $M$ and ends in $M'$.

We say that two \timed\ markings $M$, $M'$ of a given \timed\ net $N$
are \emph{\ourpradj}\ if they are bisimulation
equivalent in the induced labelled transition system.
The main topic of this paper is undecidability of the following problem:
given $N$, $M$ and $M'$, decide whether $M$ and $M'$ are \ourpradj.

The aim of performance equivalence is not only to relate functionalities of two systems,
but also their \emph{performance}, i.e.~amount of resources necessary for these functionalities.
We illustrate this aspect with a simple example.

\begin{example}
Consider a durational Petri net with the following transition rules (essentially a finite system):
\begin{align*}
p & \tranrule{a} p_a & p_a & \tranrule{b} p_{ab} & q_a & \tranrule{a} q' \\
p & \tranrule{b} p_b & p_b & \tranrule{a} p_{ab}  & q_b & \tranrule{b} q'
\end{align*}
and two its markings, $0 \tr p$ and $0 \tr q_a q_b$.
According to the standard bisimulation equivalence the ({\untiming}s of) two markings are equivalent,
as both can perform actions $a$ and $b$ in any order.
On the other hand the two marking are distinguished by performance equivalence, as they differ in performance.
For instance, the second transition in the transition sequence
$$ 0 \tr q_a q_b \trans{a, 0} (0 \tr q_b) (1 \tr q') \trans{b, 0} 1 \tr q' q' $$
can not be matched properly from $0 \tr p$, as the only possible option is
$$ 0 \tr p \trans{a, 0} 1 \tr p_a \trans{1, b} 2 \tr p_{ab}.$$

Intuitively speaking, one reason behind the distinguishing power of performance equivalence
is that it can observe (amount of) concurrency in a process.
In the example above, 
the actions $a$ and $b$ are concurrent when performed from $0 \tr q_a q_b$, but not when performed from $0 \tr p$. 
In this respect, performance equivalence resembles non-interleaving process equivalences.
In fact, in the simple subclass of communication-free Petri nets (aka Basic Parallel Processes, BPP), 
performance equivalence coincides with
distributed bisimilarity, but also with causal equivalence, history-preserving bisimilarity, and location equivalence
(see~\cite{las-mfcs} and further references therein). 

It is often not obvious how to define the non-interleaving equivalences mentioned above for classes of systems
significantly larger than the simple class of BPP (or maybe BPP extended with communication). 
For instance, history-preserving bisimilarity requires explicit notion of 
concurrency (independence) between transitions; and distributed bisimilarity requires a clear separation into
local and non-local part, which is typically impossible in Petri nets. 
In contract to this, performance equivalence is a reasonable and robust notion of equivalence
easy to define in different settings and for different process classes, as it only requires durations assigned to transitions.
\end{example}

\subsection{Reachability} \label{sec:warmup1}

It is an easy exercise to show that reachability is decidable in all the four variants.
Moreover, the \timed\ semantics makes the problem drastically easier than for the ordinary nets.

Formally, the reachability problem is formulated as follows:
given a net $N$ and two \timed\ markings $M_0$, $M$, we ask whether
$M$ is reachable from $M_0$ in $N$.
This problem is easily decidable, as it is sufficient to search, roughly speaking, through a finite part
of the induced transition system containing transitions with time label
smaller than $\maxstamp{M}$ (the other transitions need not be used if we aim at reaching exactly marking $M$).

From the point of view of complexity, a more interesting variant of reachability is obtained when the given source and destination markings
$M_0$ and $M$ are non-\timed.
We ask now whether for some markings $\bar{M}_0$ and $\bar{M}$ satisfying
\[
\untime{\bar{M}_0} = M_0 \qquad \text{and} \qquad \untime{\bar{M}} = M,
\]
$\bar{M}$ is reachable in $N$ from $\bar{M}_0$.
Hence, we specify only the number of tokens on each place in the source and destination marking, 
and leave their time-stamps unspecified.

As an easy observation, this variant of reachability is decidable in the  \localtime\ \patient\ semantics.
Indeed, \localtime\ \patient\ semantics does not differ significantly from
non-\timed\ semantics in the context of the latter variant of the reachability problem. 
Hence, a \timed\ net may be faithfully simulated by a non-\timed\ one, and therefore reachability is decidable.

On the other hand,
all the other three variants of semantics are different from  the  \localtime\ \patient~(and hence also from the non-\timed\ setting) in that
in a given reachable marking, some of the transitions may be non-fireable in the former
while fireable in the latter. That is to say, our \timed\ settings pose some additional
restriction of fireability of a transition in a marking. We leave investigation of decidability of reachability as the possible further work.

\subsection{\Patient\ semantics} \label{sec:warmup2}

It is an easy observation that \patient\ semantics can faithfully
simulate ordinary (non-\timed) semantics, and therefore we have:

\begin{proposition}
The problem of \ourpr\ is undecidable for Petri nets under (\globaltime\ or \localtime)
\patient\ semantics.
% for any fixed transition duration function.
\end{proposition}

The easy proof is  by reduction from bisimulation equivalence of ordinary Petri nets.
Given such a net $N$, we extend it by one special place $p$ and add this place both to
pre-places and post-places of any transition. In the initial marking, $p$ is marked with a single token.
Therefore during a run, place $p$ will always have precisely one token, with the ''latest'' time-stamp;
and this time-stamp will be always observable at the next transition.
Hence, \ourpr\ between so extended nets coincides with bisimulation equivalence of the
original ordinary nets.

The case of \impatient\ semantics is more difficult.
We were able to show undecidability only in the \globaltime\ variant (the whole Section~\ref{s:proof} is devoted to the undecidability proof).
Decidability of \ourpr\ in the last, \localtime\ \impatient\ variant, remains still open.

\section{Undecidability under \globaltime\ \impatient\ semantics}
\label{s:proof}

In this section, we concentrate on proving the following result:

\begin{theorem}
The problem of \ourpr\ is undecidable for Petri nets under \globaltime\ \impatient\ semantics.
% for any fixed transition cost function.
\end{theorem}

% \begin{proof}

The proof is by reduction from the (undecidable) halting problem of Minsky machines, and
occupies the rest of this section.
It is motivated by the Jan\v{c}ar's proof for ordinary Petri nets~\cite{jan};
however, a new insight was necessary to adapt this proof to the \timed\ setting.

A Minsky deterministic 2-counter machine $\M$ consists of two counters $\mathtt{c_0}, \mathtt{c_1}$ and
a set of $n$ labelled instructions

\begin{center}
\begin{tabular}{ll}
$\mathtt{1}$ : & $\mathtt{instr_1}$ \\
& \ldots \\
$\mathtt{n}$ : & $\mathtt{instr_n}$
\end{tabular}
\end{center}

\noindent
each instruction in one of the following forms:

\instrInc

\instrIf

\instrHalt

\noindent
Variables $i, j$ and $k$ range over $\{1 \ldots n\}$ and $b$ over $\{0, 1\}$.
We say that $\M$ \emph{halts} if starting from $\mathtt{1 : instr_1}$
and $\mathtt{c_0} = \mathtt{c_1} = 0$, the unique run of $\M$ ends in instruction $\mathtt{n: halt}$.

Given a  Minsky machine $\M$,
we will construct a Petri net $N_\M$ and two markings
in such a way that the markings are \ourpradj\
iff $\M$ does not halt.

\newcommand{\ii}{\text{inc}}
\newcommand{\dd}{\text{dec}}
\newcommand{\zz}{\text{zero}}

There will be places in $N_\M$ corresponding to particular instructions
of $\M$ and to particular counters.
Formally, the set of places will be $$\calP = \{ p_i, q_i, p'_i, q'_i \}_{i = 1 \ldots n} \cup
\{ b', b'', Z'_b, Z''_b \}_{b = 0,1}$$ 
(actually we only use those $p'_i$ and $q'_i$ places where $\mathtt{1 : instr_1}$ is  'zero-test or decrement' instruction)
and the labelling set
$$\calL = \{ \ii, \dd, \zz, \overline{\zz}, \tau_0, \tau_1, \omega \}.$$
Transition rules of $N_\M$ are defined as follows.
For every increment instruction we define the transition rules:

\begin{quote}
\begin{tabular}{c@{\hspace{0.3cm}}r@{\hspace{0.3cm}}c@{\hspace{0.3cm}}l@{\hspace{0.3cm}}c@{\hspace{0.3cm}}r@{\hspace{0.3cm}}c@{\hspace{0.3cm}}l}
$\mathtt{(I)}$  & $p_i$  &  $\tranrule{\ii}$  &  $p_j b' b''$
       & \hspace{4.2cm}
       & $q_i$  &  $\tranrule{\ii}$  &  $q_j b' b''$
\\
\end{tabular}
\end{quote}

\noindent
where $b \in \{ 0, 1 \}$.
For every 'zero-test or decrement' instruction we define the following transition rules:

\begin{quote}
\begin{tabular}{c@{\hspace{0.3cm}}r@{\hspace{0.3cm}}c@{\hspace{0.3cm}}l@{\hspace{0.3cm}}c@{\hspace{0.3cm}}r@{\hspace{0.3cm}}c@{\hspace{0.3cm}}l}
$\mathtt{(D)}$   & $p_i b' b''$     &  $\tranrule{\dd}$   &  $p_j$
        & \hspace{2cm}
        & $q_i b' b''$     &  $\tranrule{\dd}$   &  $q_j$
\\
$\mathtt{(Z)}$   & $p_i$            &  $\tranrule{\zz}$  &  $p_i' Z_b' Z_b''$
        & \hspace{2cm}
        & $q_i$            &  $\tranrule{\zz}$  &  $q_i' Z_b' Z_b''$
\\
$\mathtt{(Z_I)}$ & $p_i' Z_b' Z_b''$  &  $\tranrule{\overline{\zz}}$  &  $p_k$
        & \hspace{2cm}
        & $q_i' Z_b' Z_b''$  &  $\tranrule{\overline{\zz}}$  &  $q_k$
\\
$\mathtt{(Z_{II})}$ & $p_i' b'' Z_b'$    &  $\tranrule{\overline{\zz}}$  &  $p_k$
        & \hspace{2cm}
        & $q_i' b'' Z_b'$    &  $\tranrule{\overline{\zz}}$  &  $q_k$
\\
$\mathtt{(Z_{III})}$ & $p_i' b'' Z_b''$   &  $\tranrule{\overline{\zz}}$  &  $q_k$
        & \hspace{2cm}
        & $q_i' b'' Z_b''$   &  $\tranrule{\overline{\zz}}$  &  $p_k$
\\
\end{tabular}
\end{quote}

\noindent
And for the halting instruction we define:

\begin{quote}
\begin{tabular}{c@{\hspace{0.3cm}}r@{\hspace{0.3cm}}c@{\hspace{0.3cm}}l@{\hspace{0.3cm}}c@{\hspace{0.3cm}}r@{\hspace{0.3cm}}c@{\hspace{0.3cm}}l}
$\mathtt{(O)}$  & $p_n$     &  $\tranrule{\omega}$   &  $p_n$   % $\emptyset$
       & \hspace{2cm}
       &
\\
\end{tabular}
\end{quote}

\noindent
Finally, we define additional transition rules:

\begin{quote}
\begin{tabular}{c@{\hspace{0.3cm}}r@{\hspace{0.3cm}}c@{\hspace{0.6cm}}l@{\hspace{0.3cm}}c@{\hspace{0.3cm}}r@{\hspace{0.3cm}}c@{\hspace{0.3cm}}l}
$\mathtt{(T_I)}$ & $b' b''$     &  $\tranrule{\tau_b}$   &  $b' b''$
        & \hspace{2cm}
        &
\\
$\mathtt{(T_{II})}$ & $b' Z_b''$   &  $\tranrule{\tau_b}$   &  $b' b''$
        & \hspace{2cm}
        &
\\
$\mathtt{(T_{III})}$ & $b' Z_b'$    &  $\tranrule{\tau_b}$   &  $b' b''$
        & \hspace{2cm}
        &
\\
\end{tabular}
\end{quote}

\noindent
and set the duration of every transition rule to $1$.
Note that each of the rules, except $\mathtt{(O)}$, appears actually in two instances, for $b$ equal to $0$ or $1$.
We hope that it will be always clear from the context which of the two instances is considered.

As in the proof of Jan\v{c}ar \cite{jan}, we will show that machine $\M$ halts if and only if two
(singleton) \timed\ markings $0 \tr p_1$ and $0 \tr q_1$ are not \ourpradj.
% These two systems are: $(N_\M, p_0)$, which we present on the left side, and $(N_\M, q_0)$, which is presented
% on the right side.
By inspecting the rules one observes an invariant:
every marking reachable from any of the two markings contains exactly one token on
places in $\{p_i, p'_i, q_i, q'_i\}_{i=1 \ldots n}$.
% This special tokens will be called control-state tokens.
We write $(0'0'')^x$ for $x$ copies of $0' 0''$.
In the following, a marking of one of the forms (after \untiming):

$$ p_i (0' 0'')^x (1' 1'')^y  \quad \mathrm{or}\quad q_i (0' 0'')^x (1' 1'')^y$$

\noindent
will be used to
represent the machine $\M$ being in a state that enables to execute instruction $\mathtt{instr_i}$,
with the counter values $\mathtt{c_0} = x$ and $\mathtt{c_1} = y$.
In addition, the auxiliary markings of the forms (after \untiming):

$$\ \ p_i' (0' 0'')^x (1' 1'')^y,  \quad \mathrm{or}\quad   q_i' (0' 0'')^x (1' 1'')^y $$

\noindent
will be used as intermediate ones in the faithful simulation of zero tests of $\M$.
Note that not all reachable markings are of this forms, but every marking has a maximal sub-marking of this form,
and it is this sub-marking that we use to determine the state of machine $\M$.

According to the \globaltime\ semantics, a transition rule may be fired with time label $t \in \nat$
only when no transition rule is fireable with time label smaller than $t$.
Consider now any execution (a sequence of transitions) of $N_\M$ starting from
$0 \tr p_1$ or $0 \tr q_1$, and the first transition $M \trans{a,t} M'$ with time label $t$ during this execution.
All the tokens in $M$ with time-stamp smaller than $t$
are therefore not able to engage in a transition from $M$.
We say these tokens are \emph{dead} in $M$; formally, a token is dead in a marking $M$
if some transition is fireable in $M$ with time label bigger than the time-stamp of this token.
Note that a token which is dead in some $M$, remains
dead in each marking reachable from $M$,   since the semantics is \impatient.

We make the following observation:
\begin{claim}
For every $t \in \nat$, every execution of $N_\M$ from $0 \tr p_1$ or $0 \tr q_1$
contains a unique marking such that the time-stamps of all non-dead tokens are equal to $t$
(such a marking is called \emph{$t$-\equi}).
\end{claim}
Indeed, if a transition with time label $t$ appears in the execution, then the source marking $M$ of
the first such transition satisfies
the condition ($M$ contains no tokens with time-stamp larger than $t$ as
the duration of each transition rule is $1$).
Or no transition with time label $t$ appears at all, which implies that the execution is finite
and ends in a deadlock marking, in which all tokens are dead and have time-stamps less or equal $t$.
It clearly satisfies the condition of the claim.

\removed{
consider the first transition with time-stamp $t$ in a given
execution of $N_\M$.
There are two possible cases:

\begin{enumerate}
\item No such transition takes place during the execution. Then, no transition with time-stamp greater than $t$
takes place either, as in the initial marking all tokens have time-stamps $0$.
Hence, a marking is reached in which no transition is fireable, i.e., all tokens are dead and thus almost all tokens
have time-stamp $t$.
\item Otherwise, consider the marking at which this transition executes (we call this marking \emph{$t$-\equi}):
in this marking almost all tokens have time-stamp $t$.
Indeed, as this is the \emph{first} transition with time-stamp $t$, this marking cannot contain tokens with time-stamps
greater than $t$ (cost of every transition is $1$). On the other hand, if the marking would contain tokens with time-stamps
smaller than $t$ which are not dead, then the transition with time-stamp $t$ would not be fireable.
\end{enumerate}
} % reomovd

All transitions executed between two consecutive \equi{s} ($t$- and $(t{+}1)$-\equi, for some $t$)
we call a \emph{large step}.

\paragraph{}\

Now we are prepared for the reduction.
As the first case, assume that machine $\M$ halts; we will show a winning strategy for \A.
This strategy corresponds to the actual execution of $\M$, hence we call it a \emph{correct simulation}.
An execution of each instruction of the machine is represented (simulated) by one or two large steps.

A position $(M, M')$ in a play is \emph{conforming} if $M$ and $M'$ are identical, except that one of them
may have a token $t \tr p_i$ and the other $t \tr q_i$,
or one of  $M, M'$ may have $t \tr p'_i$ while the other $t \tr q'_i$.
The starting position $(0 \tr p_1, 0 \tr q_1)$ is conforming.
Consider a conforming pair of $t$-\equi{s}:
\begin{longtable}{@{}p{\tssize}|p{\tssize}}
\hfil $t \tr p_i (0' 0'')^x (1' 1'')^y$
\hfil &
\hfil $t \tr q_i (0' 0'')^x (1' 1'')^y$
\hfil \\
\end{longtable}
%We write $(0'0'')^x$ for $x$ copies of $0' 0''$.
As mentioned above, when we write $t \tr p_i (0' 0'')^x (1' 1'')^y$ we mean that time-stamps of all tokens
are $t$.
In the sequel, to succinctly write markings containing tokens with different time-stamps,
we will use a union operation on multisets, denoted by $\comp$.
\noindent
We will show how \A\ can enforce a conforming pair of $(t{+}1)$-\equi{s}\ during the play,
at the same time simulating the execution of $\M$.
Assume, without loss of generality, that $\mathtt{instr_i}$ acts on counter $\mathtt{c_0}$.
We consider all possible types of instruction $\mathtt{instr_i}$.
In each case, \A\ plays on the left-hand side and
\D\ is constantly forced to copy \A's moves on the other side.

%\begin{itemize}
%    \item[$(a)$]
\paragraph{\bf \typeInc}

    \A\ executes transition $\mathtt{(I)}$, to which \D\ must respond with $\mathtt{(I)}$.
Next, \A\ fires $x {+} y$ times transition rule $\mathtt{(T_I)}$ (first all $\tau_0$, then all $\tau_1$);
for future reference, we call such a sequence of $\mathtt{(T_I)}$ transitions a
default \emph{step completion}.
During the completion, the responses of \D\ are uniquely determined,
as transitions rules $\mathtt{(T_{II})}$ and $\mathtt{(T_{III})}$ are not fireable.

\begin{longtable}{@{}p{\tssize}|p{\tssize}}
\hfil $t \tr p_i (0' 0'')^x (1' 1'')^y$
\hfil &
\hfil $t \tr q_i (0' 0'')^x (1' 1'')^y$
\hfil \\
\hfil \hspace{0.5cm} $\dtrans{\mathtt{(I)}}$
\hfil &
\hfil \hspace{0.5cm} $\dtrans{\mathtt{(I)}}$
\hfil \\
\hfil $(t{+}1) \tr p_j (0' 0'') \comp t \tr (0' 0'')^x (1' 1'')^y$
\hfil &
\hfil $(t{+}1) \tr q_j (0' 0'') \comp t \tr (0' 0'')^x (1' 1'')^y$
\hfil \\
\hfil \hspace{0.5cm} $\dtrans{\mathtt{(T_I)}^x}$
\hfil &
\hfil \hspace{0.5cm} $\dtrans{\mathtt{(T_I)}^x}$
\hfil \\
\hfil $(t{+}1) \tr p_j (0' 0'')^{x+1} \comp t \tr (1' 1'')^y$
\hfil &
\hfil $(t{+}1) \tr q_j (0' 0'')^{x+1} \comp t \tr (1' 1'')^y$
\hfil \\
\hfil \hspace{0.5cm} $\dtrans{\mathtt{(T_I)}^y}$
\hfil &
\hfil \hspace{0.5cm} $\dtrans{\mathtt{(T_I)}^y}$
\hfil \\
\hfil $(t{+}1) \tr p_j (0' 0'')^{x+1} (1' 1'')^y$
\hfil &
\hfil $(t{+}1) \tr q_j (0' 0'')^{x+1} (1' 1'')^y$
\hfil \\
\end{longtable}

For convenience, we decorate each transition
by the identifier of the relevant transition rule, in place of its label.

Both the resulting markings are $(t{+}1)$-\equi{s} which
completes the simulation of instruction $\mathtt{instr_i}$.
They represent the machine $\M$ in a state that enables to execute instruction $\mathtt{instr_j}$,
and the counters values are as expected.

\paragraph{\bf \typeIf}
%    \item[$(b)$]

    In this case, the behaviour of \A\ depends on the value of counter $\mathtt{c_0}$, i.e.
the value of $x$.

    If $x > 0$,  \A\ start with transition $\mathtt{(D)}$ and completes the step in the default manner:
\begin{longtable}{@{}p{\tssize}|p{\tssize}}
\hfil $t \tr p_i (0' 0'')^x (1' 1'')^y$
\hfil &
\hfil $t \tr q_i (0' 0'')^x (1' 1'')^y$
\hfil \\
\hfil \hspace{0.5cm} $\dtrans{\mathtt{(D)}}$
\hfil &
\hfil \hspace{0.5cm} $\dtrans{\mathtt{(D)}}$
\hfil \\
\hfil $(t{+}1) \tr p_j \comp t \tr (0' 0'')^{x-1} (1' 1'')^y$
\hfil &
\hfil $(t{+}1) \tr q_j \comp t \tr (0' 0'')^{x-1} (1' 1'')^y$
\hfil \\
\hfil \hspace{1.4cm} $\dtrans{\mathtt{(T_I)}^{x+y-1}}$
\hfil &
\hfil \hspace{1.4cm} $\dtrans{\mathtt{(T_I)}^{x+y-1}}$
\hfil \\
\hfil $(t{+}1) \tr p_j (0' 0'')^{x-1} (1' 1'')^y$
\hfil &
\hfil $(t{+}1) \tr q_j (0' 0'')^{x-1} (1' 1'')^y$
\hfil \\
\end{longtable}

        In the other case, when $x = 0$, \A\ executes transition $\mathtt{(Z)}$, to which \D\ must respond with $\mathtt{(Z)}$.
Then \A\ completes the step in the default manner.

\begin{longtable}{@{}p{\tssize}|p{\tssize}}
\hfil $t \tr p_i (1' 1'')^y$
\hfil &
\hfil $t \tr q_i (1' 1'')^y$
\hfil \\
\hfil \hspace{0.5cm} $\dtrans{\mathtt{(Z)}}$
\hfil &
\hfil \hspace{0.5cm} $\dtrans{\mathtt{(Z)}}$
\hfil \\
\hfil $(t{+}1) \tr p_i' (Z_0' Z_0'') \comp t \tr (1' 1'')^y$
\hfil &
\hfil $(t{+}1) \tr q_i' (Z_0' Z_0'') \comp t \tr (1' 1'')^y$
\hfil \\
\hfil \hspace{0.75cm} $\dtrans{\mathtt{(T_I)}^y}$
\hfil &
\hfil \hspace{0.75cm} $\dtrans{\mathtt{(T_I)}^y}$
\hfil \\
\hfil $(t{+}1) \tr p_i' (Z_0' Z_0'') (1' 1'')^y$
\hfil &
\hfil $(t{+}1) \tr q_i' (Z_0' Z_0'') (1' 1'')^y$
\hfil \\
\end{longtable}

\noindent
    Then \A\ starts the next large step with transition $\mathtt{(Z_I)}$. 
As there are no $0''$ tokens in the marking, transitions $\mathtt{(Z_{II})}$ and $\mathtt{(Z_{III})}$ are not 
fireable and \D\ must respond with $\mathtt{(Z_I)}$. Next, \A\ completes the step in the default manner, 
reaching $(t{+}2)$-\equi{s}, thus finalising the simulation of instruction $\mathtt{instr_i}$ as expected.

\begin{longtable}{@{}p{\tssize}|p{\tssize}}
\hfil $(t{+}1) \tr p_i' (Z_0' Z_0'') (1' 1'')^y$
\hfil &
\hfil $(t{+}1) \tr q_i' (Z_0' Z_0'') (1' 1'')^y$
\hfil \\
\hfil \hspace{0.5cm} $\dtrans{\mathtt{(Z_I)}}$
\hfil &
\hfil \hspace{0.5cm} $\dtrans{\mathtt{(Z_I)}}$
\hfil \\
\hfil $(t{+}2) \tr p_k \comp (t{+}1) \tr (1' 1'')^y$
\hfil &
\hfil $(t{+}2) \tr q_k \comp (t{+}1) \tr (1' 1'')^y$
\hfil \\
\hfil \hspace{0.65cm} $\dtrans{\mathtt{(T_I)}^y}$
\hfil &
\hfil \hspace{0.65cm} $\dtrans{\mathtt{(T_I)}^y}$
\hfil \\
\hfil $(t{+}2) \tr p_k (1' 1'')^y$
\hfil &
\hfil $(t{+}2) \tr q_k (1' 1'')^y$
\hfil \\
\end{longtable}

    % This finalises the simulation of instruction $\mathtt{i}$.

    In both cases the final position is a conforming pair of \equi{s}.

%    \item[$(c)$]
\paragraph{\bf \typeHalt}

    Player \A\ executes transition $\mathtt{(O)}$; \D\ has no response and losses the game.
% \end{itemize}

\paragraph{}\

As the second case, assume that machine $\M$ does not halt; we will show a wining strategy for \D.
As long as \A\ plays a correct simulation, \D's behaviour is determined.
However, a position enabling firing of transition rule $\mathtt{(O)}$
% representing execution of instruction $\mathtt{n: halt}$
is never reached, and the game never ends -- \D\ wins.

\A\ can diverge from the correct simulation (\emph{cheat}) in two ways: \emph{insignificant} or \emph{significant}.
The former means that the \equi{s}\ are the same as in the correct simulation.
This kind of cheating involves shuffling the order of transitions within one large step and changing the side
on which \A\ is playing.
With such behaviour, he still essentially simulates the execution of $\M$:
he will never fire transition rule $\mathtt{(O)}$,
% cannot reach instruction $\mathtt{n: halt}$,
thus \D\ wins.

\A\ can also cheat in a significant manner, simulating a positive zero test, when it should decrease the counter.
However, \D\ can respond to such behaviour in a way which will make the next \equi{s}\ identical, modulo dead tokens.
Afterwards, \D\ can copy \A's behaviour exactly, and hence win the game.

In what follows, we give an exhaustive overview of possible \A's cheating behaviour, and show how \D\ can 
respond to them to ensure her win.
As before, we start with a conforming \equi\ position and we assume that instruction
$\mathtt{instr_i}$ uses counter $\mathtt{c_0}$.

\begin{longtable}{@{}p{\tssize}|p{\tssize}}
\hfil $t \tr p_i (0' 0'')^x (1' 1'')^y$
\hfil &
\hfil $t \tr q_i (0' 0'')^x (1' 1'')^y$
\hfil \\
\end{longtable}

\noindent
We consider below all possible types of instruction $\mathtt{instr_i}$.
In each case we assume that the first move of \A\ is on the left-hand side.

%\begin{itemize}
%\item[$(a)$]

\paragraph{\bf \typeInc}
    \A\ can, in any order, execute transitions $\mathtt{(I)}$ and $\mathtt{(T_I)}$, which \D\ must (and can) copy.
    The \equi{s}\ reached in the end cannot be different then those appearing in the correct simulation.
    It is hence an insignificant cheating.

%\item[$(b)$]
\paragraph{\bf \typeIf}
    If counter $\mathtt{c_0}$ used by the instruction is equal to 0, \A\ can only cheat insignificantly.
    During the first large step involved in the simulation, \A\
    can choose an arbitrary ordering of $\mathtt{(Z)}$ and $\mathtt{(T_I)}$ transitions, as
    it does not influence the resulting \equi{s}:

\begin{longtable}{@{}p{\tssize}|p{\tssize}}
\hfil $(t{+}1) \tr p_i' (Z_0' Z_0'') (1' 1'')^y$
\hfil &
\hfil $(t{+}1) \tr q_i' (Z_0' Z_0'') (1' 1'')^y$
\hfil \\
\end{longtable}

\noindent
    In the second large step, \A's choice is again limited to shuffling of
$\mathtt{(Z_I)}$ and $\mathtt{(T_I)}$ transitions,
    as none of the $\mathtt{(Z_{II})}$, $\mathtt{(Z_{III})}$, $\mathtt{(T_{II})}$ or $\mathtt{(T_{III})}$ transition rules is fireable.
    It has no influence on the resulting \equi{s}, which completes the simulation of $\mathtt{instr_i}$:

\begin{longtable}{@{}p{\tssize}|p{\tssize}}
\hfil $(t{+}2) \tr p_k (1' 1'')^y$
\hfil &
\hfil $(t{+}2) \tr q_k (1' 1'')^y$
\hfil \\
\end{longtable}

    Now assume that $\mathtt{c_0}$ is non-zero, $x > 0$. \A\ can cheat insignificantly,
    if he decides to decrease the counter.
    As before, he will be limited to shuffling around the $\mathtt{(D)}$ and $\mathtt{(T_I)}$ transitions,
    with no influence on the resulting \equi{s}:

\begin{longtable}{@{}p{\tssize}|p{\tssize}}
\hfil $(t{+}1) \tr p_j (0' 0'')^{x-1} (1' 1'')^y$
\hfil &
\hfil $(t{+}1) \tr q_j (0' 0'')^{x-1} (1' 1'')^y$
\hfil \\
\end{longtable}

    In all the cases above, \D\ was forced to copy the moves of \A.
    A more interesting case is when \A\ decides to cheat significantly, and executes
    transition $\mathtt{(Z)}$ despite that $x > 0$.
    \D\ is forced to respond with $\mathtt{(Z)}$.
    Then, after completion of the step, irrespectively of the shuffling of the
    $\mathtt{(T_I)}$ and $\mathtt{(Z)}$ transitions (no other transition rules are fireable),
    the \equi{s}\ will have the from:

\begin{longtable}{@{}p{\tssize}|p{\tssize}}
\hfil $(t{+}1) \tr p_i' (Z_0' Z_0'') (0' 0'')^x (1' 1'')^y$
\hfil &
\hfil $(t{+}1) \tr q_i' (Z_0' Z_0'') (0' 0'')^x (1' 1'')^y$
\hfil \\
\end{longtable}

\noindent
    Note that the only fireable $\tau_1$-labeled transition rule is $\mathtt{(T_I)}$, activated by
    tokens $(t{+}1) \tr 1'$ and $(t{+}1) \tr 1''$
    (as there are no $Z_1'$ or $Z_1''$ tokens).
    It does not matter for the next $(t+2)$-\equi{s}\
    how firings of $\mathtt{(T_I)}$ are interleaved with the other transitions.
    Hence, we omit it in the presentation below and proceed as if we were in the position:

\begin{longtable}{@{}p{\tssize}|p{\tssize}}
\hfil $(t{+}1) \tr p_i' (Z_0' Z_0'') (0' 0'')^x$
\hfil & \hspace{0.5cm}
\hfil $(t{+}1) \tr q_i' (Z_0' Z_0'') (0' 0'')^x$
\hfil \hspace{0.7cm} ($\dagger$)\\
\end{longtable}

    In all of the cases below (except for one, which is explained separately), \D\ will respond in such a way that
    the resulting \equi{s}\ are identical.
    From that point on she may copy exactly the \A's\ moves and hence win the game.

\begin{enumerate}
%:
    \item
    If \A\ executes $\mathtt{(Z_I)}$, \D\ responds with $\mathtt{(Z_{III})}$.
    The step can be completed only by $x$ executions of the $\mathtt{(T_I)}$ transition on the left
    and $x{-}1$ executions
    of $\mathtt{(T_I)}$ and one execution of $\mathtt{(T_{III})}$ on the right.
    These transitions have identical labels, hence it is irrelevant in what order they are executed
    and on which side \A\ is playing.

\begin{longtable}{@{}p{\tssize}|p{\tssize}}
\hfil $(t{+}1) \tr p_i' (Z_0' Z_0'') (0' 0'')^x$
\hfil &
\hfil $(t{+}1) \tr q_i' (Z_0' Z_0'') (0' 0'')^x$
\hfil \\
\hfil \hspace{0.5cm} $\dtrans{\mathtt{(Z_I)}}$
\hfil &
\hfil \hspace{0.5cm} $\dtrans{\mathtt{(Z_{III})}}$
\hfil \\
\hfil $(t{+}2) \tr p_k \comp (t{+}1) \tr (0' 0'')^{x}$
\hfil &
\hfil $(t{+}2) \tr p_k \comp (t{+}1) \tr (0' Z_0') (0' 0'')^{x-1}$
\hfil \\
\hfil \hspace{0.65cm} $\dtrans{\mathtt{(T_I)}^x}$
\hfil &
\hfil \hspace{2.5cm} $\dtrans{\mathtt{(T_I)}^u\mathtt{(T_{III})}\mathtt{(T_I)}^{x-u-1}}$
\hfil \\
\hfil $(t + 2) \tr p_k (0' 0'')^x$
\hfil &
\hfil $(t + 2) \tr p_k (0' 0'')^x$
\hfil \\
\end{longtable}

    \item
    If \A\ executes transition $\mathtt{(Z_{II})}$, \D\ responds with $\mathtt{(Z_{III})}$.
    Completing the step involves $x{-}1$ firings of $\mathtt{(T_I)}$ and one firing of $\mathtt{(T_{II})}$ on the left,
    and $x{-}1$ firings of $\mathtt{(T_I)}$ and one firing of $\mathtt{(T_{III})}$ on the right.
    As above, the \A's\ choices at this stage have no influence on the resulting \equi{s}.

\begin{longtable}{@{}p{\tssize}|p{\tssize}}
\hfil $(t{+}1) \tr p_i' (Z_0' Z_0'') (0' 0'')^x$
\hfil &
\hfil $(t{+}1) \tr q_i' (Z_0' Z_0'') (0' 0'')^x$
\hfil \\
\hfil \hspace{0.5cm} $\dtrans{\mathtt{(Z_{II})}}$
\hfil &
\hfil \hspace{0.5cm} $\dtrans{\mathtt{(Z_{III})}}$
\hfil \\
\hfil $(t{+}2) \tr p_k \comp (t{+}1) \tr (0' Z_0'') (0' 0'')^{x-1}$
\hfil &
\hfil $(t{+}2) \tr p_k \comp (t{+}1) \tr (0' Z_0') (0' 0'')^{x-1}$
\hfil \\
\hfil \hspace{2.5cm} $\dtrans{\mathtt{(T_I)}^u\mathtt{(T_{II})}\mathtt{(T_I)}^{x-u-1}}$
\hfil &
\hfil \hspace{2.5cm} $\dtrans{\mathtt{(T_I)}^u\mathtt{(T_{III})}\mathtt{(T_I)}^{x-u-1}}$
\hfil \\
\hfil $(t {+} 2) \tr p_k (0' 0'')^x$
\hfil &
\hfil $(t {+} 2) \tr p_k (0' 0'')^x$
\hfil \\
\end{longtable}

    \item
    If \A\ executes transition $\mathtt{(Z_{III})}$, \D\ responds with $\mathtt{(Z_{II})}$. This case is similar to the one above.

  \item
  If \A\ executes transition $\mathtt{(T_I)}$, \D\ responds depending on the value of $x$. If $x > 1$,
  she executes $\mathtt{(T_I)}$.
  Notice that the tokens $(t{+}2) \tr (0' 0'')$ have no influence on the $(t{+}1)$-time-stamped tokens.
  Hence \D\ can continue as if the configuration was ($\dagger$), but with one less $0'$ and $0''$ tokens.

\begin{longtable}{@{}p{\tssize}|p{\tssize}}
\hfil $(t{+}1) \tr p_i' (Z_0' Z_0'') (0' 0'')^x$
\hfil &
\hfil $(t{+}1) \tr q_i' (Z_0' Z_0'') (0' 0'')^x$
\hfil \\
\hfil \hspace{0.5cm} $\dtrans{\mathtt{(T_I)}}$
\hfil &
\hfil \hspace{0.5cm} $\dtrans{\mathtt{(T_I)}}$
\hfil \\
\hfil $(t{+}1) \tr p_i' (Z_0' Z_0'') (0' 0'')^{x-1} \comp$ 
\hfil &
\hfil $(t{+}1) \tr q_i' (Z_0' Z_0'') (0' 0'')^{x-1} \comp$ 
\hfil \\
\hfil $(t{+}2) \tr (0' 0'')$ \hfil & \hfil  $(t{+}2) \tr (0' 0'')$ \hfil
\end{longtable}

    If $x=1$, \D\ executes $\mathtt{(T_{III})}$.
Completing the step involves firing of $\mathtt{(Z_I)}$ on the left and
$\mathtt{(Z_{III})}$ on the right.

\begin{longtable}{@{}p{\tssize}|p{\tssize}}
\hfil $(t{+}1) \tr p_i' (Z_0' Z_0'') (0' 0'')$
\hfil &
\hfil $(t{+}1) \tr q_i' (Z_0' Z_0'') (0' 0'')$
\hfil \\
\hfil \hspace{0.5cm} $\dtrans{\mathtt{(T_I)}}$
\hfil &
\hfil \hspace{0.5cm} $\dtrans{\mathtt{(T_{III})}}$
\hfil \\
\hfil $(t{+}1) \tr p_i' (Z_0' Z_0'') \comp (t{+}2) \tr (0' 0'')$
\hfil &
\hfil $(t{+}1) \tr q_i' (0'' Z_0'') \comp (t{+}2) \tr (0' 0'')$
\hfil \\
\hfil \hspace{0.5cm} $\dtrans{\mathtt{(Z_I)}}$
\hfil &
\hfil \hspace{0.5cm} $\dtrans{\mathtt{(Z_{III})}}$
\hfil \\
\hfil $(t {+} 2) \tr p_k (0' 0'')$
\hfil &
\hfil $(t {+} 2) \tr p_k (0' 0'')$
\hfil \\
\end{longtable}

  \item If \A\ executes $\mathtt{(T_{II})}$, \D\ responds with $\mathtt{(T_{III})}$.

\begin{longtable}{@{}p{\tssize}|p{\tssize}}
\hfil $(t{+}1) \tr p_i' (Z_0' Z_0'') (0' 0'')^x$
\hfil &
\hfil $(t{+}1) \tr q_i' (Z_0' Z_0'') (0' 0'')^x$
\hfil \\
\hfil \hspace{0.5cm} $\dtrans{\mathtt{(T_{II})}}$
\hfil &
\hfil \hspace{0.5cm} $\dtrans{\mathtt{(T_{III})}}$
\hfil \\
\hfil $(t{+}1) \tr p_i' (0'' Z_0') (0' 0'')^{x-1} \comp (t{+}2) \tr (0' 0'')$
\hfil &
\hfil $(t{+}1) \tr q_i' (0'' Z_0'') (0' 0'')^{x-1} \comp (t{+}2) \tr (0' 0'')$
\hfil \\
\end{longtable}

    Now the play can continue in one of two ways, depending on the behaviour of \A. 
    Before it is decided, as long as \A\ executes $\mathtt{(T_I)}$, \D\ responds with $\mathtt{(T_I)}$:

\begin{longtable}{@{}p{\tssize}|p{\tssize}}
\hfil $(t{+}1) \tr p_i' (0'' Z_0') (0' 0'')^{x-1} \comp (t{+}2) \tr (0' 0'')$
\hfil &
\hfil $(t{+}1) \tr q_i' (0'' Z_0'') (0' 0'')^{x-1} \comp (t{+}2) \tr (0' 0'')$
\hfil \\
\hfil \hspace{0.65cm} $\dtrans{\mathtt{(T_I)}^u}$
\hfil &
\hfil \hspace{0.65cm} $\dtrans{\mathtt{(T_I)}^u}$
\hfil \\
\hfil $(t{+}1) \tr p_i' (0'' Z_0') (0' 0'')^{x-u-1} \comp$ 
\hfil &
\hfil $(t{+}1) \tr q_i' (0'' Z_0'') (0' 0'')^{x-u-1} \comp$ 
\hfil \\
\hfil $(t{+}2) \tr (0' 0'')^{u+1}$ \hfil & \hfil  $(t{+}2) \tr (0' 0'')^{u+1}$
\end{longtable}

At some point \A\ must decide between executing, on the left-hand side, of $\mathtt{(Z_{II})}$ (equivalently, $\mathtt{(Z_{III})}$ on the right) or
$\mathtt{(T_{III})}$ (equivalently, $\mathtt{(T_{II})}$ on the right). In the first case, \D\ responds as follows:

\begin{longtable}{@{}p{\tssize}|p{\tssize}}
\hfil $(t{+}1) \tr p_i' (0'' Z_0') (0' 0'')^{x-u-1} \comp$ 
\hfil &
\hfil $(t{+}1) \tr q_i' (0'' Z_0'') (0' 0'')^{x-u-1} \comp$ 
\hfil \\
\hfil $(t{+}2) \tr (0' 0'')^{u+1}$ \hfil & \hfil $(t{+}2) \tr (0' 0'')^{u+1}$\\
\hfil \hspace{0.5cm} $\dtrans{\mathtt{(Z_{II})}}$
\hfil &
\hfil \hspace{0.5cm} $\dtrans{\mathtt{(Z_{III})}}$
\hfil \\
\hfil $(t{+}1) \tr (0' 0'')^{x-u-1} \comp (t{+}2) \tr p_k (0' 0'')^{u+1}$
\hfil &
\hfil $(t{+}1) \tr (0' 0'')^{x-u-1} \comp (t{+}2) \tr p_k (0' 0'')^{u+1}$
\hfil \\
\hfil \hspace{1.3cm} $\dtrans{\mathtt{(T_I)}^{x-u-1}}$
\hfil &
\hfil \hspace{1.3cm} $\dtrans{\mathtt{(T_I)}^{x-u-1}}$
\hfil \\
\hfil $(t {+} 2) \tr p_k (0' 0'')^x$
\hfil &
\hfil $(t {+} 2) \tr p_k (0' 0'')^x$
\hfil \\
\end{longtable}

\noindent
    whereas in the latter case:

\begin{longtable}{@{}p{\tssize}|p{\tssize}}
\hfil $(t{+}1) \tr p_i' (0'' Z_0') (0' 0'')^{x-u-1} \comp$ 
\hfil &
\hfil $(t{+}1) \tr q_i' (0'' Z_0'') (0' 0'')^{x-u-1} \comp$ 
\hfil \\
\hfil $(t{+}2) \tr (0' 0'')^{u+1}$ \hfil & \hfil $(t{+}2) \tr (0' 0'')^{u+1}$ \\
\hfil \hspace{0.5cm} $\dtrans{\mathtt{(T_{III})}}$
\hfil &
\hfil \hspace{0.5cm} $\dtrans{\mathtt{(T_{II})}}$
\hfil \\
\hfil $(t{+}1) \tr p_i' 0'' 0'' (0' 0'')^{x-u-2} \comp (t{+}2) \tr (0' 0'')^{u+2}$
\hfil &
\hfil $(t{+}1) \tr q_i' 0'' 0'' (0' 0'')^{x-u-2} \comp (t{+}2) \tr (0' 0'')^{u+2}$
\hfil \\
\hfil \hspace{1.3cm} $\dtrans{\mathtt{(T_I)}^{x-u-2}}$
\hfil &
\hfil \hspace{1.3cm} $\dtrans{\mathtt{(T_I)}^{x-u-2}}$
\hfil \\
\hfil $(t{+}1) \tr p_i' 0'' 0'' \comp (t{+}2) \tr (0' 0'')^x$
\hfil &
\hfil $(t{+}1) \tr q_i' 0'' 0'' \comp (t{+}2) \tr (0' 0'')^x$
\hfil \\
\end{longtable}

    This is the only case so far in which the resulting markings are not identical.
    However, the $(t{+}1)$-time-stamped tokens on both sides are dead, whereas the $(t{+}2)$-time-stamped
    ones are identical.
    Hence, \D\ can still copy in future all the transitions that \A\ makes,
    even if it the configurations do not represent a  proper machine state anymore.

    \item
    If \A\ executes $\mathtt{(T_{III})}$, \D\ responds with $\mathtt{(T_{II})}$. What follows is
    similar to the previous point.

\end{enumerate}

% \item[$(c)$]
\paragraph{\bf \typeHalt}
    Simulation of instruction $\mathtt{n : halt}$ may happen only if \A\ have cheated earlier
    in a significant manner.
    As we have shown above, in such a case the markings on both sides are identical and \D\ still wins.

%\end{itemize}

\paragraph{}\

We have shown that the initial markings $0 \tr p_1$ and $0 \tr q_1$ are \ourpradj\ if and only if machine  $\M$ does not halt.
Hence \ourpr\ is undecidable.

% \end{proof}

%\section{Decidability of the reachability problem}
%\label{s:reach}

\section{Concluding remarks}
\label{s:concl}

We have investigated Petri nets under four different durational semantics, and the corresponding
variant of bisimulation equivalence, called performance equivalence.
We have proved that in three of the four cases, performance
equivalence is undecidable.
Hence, unfortunately, our results do not confirm a conjecture that performance equivalence
might be easier to decide than classical bisimulation equivalence.
However in the fourth variant, i.e., \localtime\ \impatient\ semantics,
we were able neither to confirm nor to falsify the conjecture.
There is hence still a hope that this last variant might admit an effective decision procedure
and we conjecture that this is really the case. Verifying this conjecture is the main open problem left.

As an encouraging observation we mention a partial decomposition property which holds in
case of \localtime\ \impatient\ semantics.
Consider a pair of performance equivalent markings, $t \tr M$ and $t \tr M'$, and some transition
of one of them $t \tr M \trans{a, t} t \tr M_1 \comp (t{+}1) \tr M_2$, assuming $\dur{a} = 1$,
which may be answered by some $t \tr M' \trans{a,t} t \tr M'_1 \comp (t{+}1) \tr M'_2$.
An easy observation is that $(t{+}1) \tr M_2$ and $(t{+}1) \tr M'_2$ must be necessarily
equivalent, as Spoiler may decide to use only time labels greater than $t$ from this point on.
More generally, if $M$ and $M'$ are equivalent, than for any $t$, their submarkings
containing tokens with time-stamps greater than $t$ are equivalent too.

The durational nets may be considered as a special case of \emph{data nets}~\cite{datanets}, once one slightly relaxes this
notion by allowing data domain different than equality data $(\mathbb{N}, =)$ or ordered data $(\mathbb{Q}, \leq)$. 
Fix a logical structure called \emph{data domain}.
Data Petri nets are very much like ordinary Petri nets,
with a proviso that every token carries a data value (an element of data domain), 
and every transition rule has a constraint on data values of tokens involved.
The transition constraints are specified by formulas
in the language of data domain; for our purposes, it is enough to restrict to quantifier-free formulas. 

Within this generic setting, durational Petri nets can be seen as a syntactic 
restriction of data nets over data domain $(\mathbb{Z}, +1, =)$, i.e.~integers with the successor function $+1$ and equality. 
In particular, transition constraints of durational nets are specified using equality and +1.
In this perspective, performance equivalence is exactly the natural interpretation of the standard bisimulation equivalence
in data nets. As data nets always generalize ordinary nets, the bisimulation equivalence will be always undecidable in general.
The present paper may be therefore seen as a search for interesting syntactic restrictions that would guarantee decidability
of bisimulation equivalence for data nets over data domain $(\mathbb{Z}, +1, =)$.

For the sake of comparision, if data universe is chosen to be $(\mathbb{Z}, +1, \leq)$ then data nets generalize discrete-timed Petri nets;
and for data universe $(\mathbb{Q}, +1, \leq)$ we obtain a generalization of dense-timed Petri nets. 

Now we turn to the reachability problem.
In Section~\ref{s:prel} we have identified two different variants of this problem: one in which source and destination markings are specified exactly
(easy to solve in all the four variants of semantics), and another one in which source and destination markings are given as non-\timed\ ones.
While the latter variant of the problem is easily decidable for \localtime\ \patient\ semantics, we do not know its status for the other variants. 
Investigation of this question is a natural continuation of the work presented in this paper.
We suspect that there may be a reduction from \timed\ nets to nets with hierarchical inhibitor arcs, a model with decidable reachability 
problem according to~\cite{Rein} (see also~\cite{Bonnet11} for a self-contained proof of reachabilty for nets with one zero test).

\paragraph{Acknowledgements.} We are very grateful to the anonymous reviewers for careful reading of the draft and
many valuable comments and suggestions.

\end{document}